\documentclass[12pt]{article}
\usepackage[utf8]{inputenc}
\usepackage[T1]{fontenc}
\usepackage{subfigure}
\usepackage{graphicx}
\usepackage[doublespacing]{setspace}
\usepackage{fullpage}
\usepackage{subfigure}

\title{Evaluation of the training objectives with surface electromyography}
\author{Paulina Trybek$^{1}$, Michał Nowakowski$^{2}$ \and Łukasz Machura$^{1}$ \\
\mbox{}\\
$^1$Divison of Computational Physics and Electronics\\
Silesian Center for Education and Interdisciplinary Research\\Institute of Physics, University of Silesia, Katowice, Poland\\
$^2$Jagiellonian University School of Medicine, Krak{\'o}w, Poland\\
p.trybek@us.edu.pl}

\begin{document}
\maketitle

\begin{abstract}
In this work the multifractal analysis of the kinesiological surface electromyographic 
signal is proposed. The goal was to investigate the level of neuromuscular activation 
during complex movements on the laparoscopic trainer. The basic issue of this work 
concerns the changes observed in the signal obtained from the complete beginner in the 
field of using laparoscopic tools and the same person subjected to the series of 
training. To quantify the complexity of the kinesiological sEMG, the nonlinear analysis 
technique, namely the MultiFractal Detrended Fluctuation Analysis was adopted. The analysis 
was based on the parameters describing the multifractal spectrum -- Hurst exponent and 
the spectrum width. The statistically significant differences for a selected group of 
muscles at the different states (before and after training) are presented. Additionally, 
as the base case, the relaxation state was considered and compared with the working states. 
\end{abstract}

\section{Introduction}
The improper patterns of muscle recruitment are often the cause for the decrease of the 
efficiency of the movements in many aspects of life. This automatically entails reduction 
of the precision and is also the reason for an increasing fatigue 
\cite{aggarwal2004laparoscopic,forsman2001motor}. This work concerns the issue of 
the ergonomic handling of the laparoscopic instruments. At the moment most of the assessments are 
performed subjectively by a trainer either locally or - in some rare cases remotely - with 
the use of the video-assessment tools. These methods lack both repeatability and specificity. 
The effectiveness of the surface electromyography (sEMG) in the assessment of the level of 
involvement of the muscular system has some strong evidence 
\cite{hug2011can,merletti2004electromyography,barbero2012atlas}. Yet still some relevant 
and unsolved issues need to be addressed. To compare the signal of a muscle activation at 
the different levels of involvement for the selected individual muscle groups located in the 
human arms a simple experiment was invented. As its details will be presented in the next 
section, here we will only mention that we have recorded the sEMG signals from the untrained 
and trained volunteers. The detected differences in the signals received from the volunteers 
performing complex movements on laparoscopic trainer can open new opportunities of using sEMG 
as a helpful tool in the process of the individual training in rather difficult motor tasks.
Several works documenting the physiological phenomena and focusing on the nonlinear dynamics 
including the chaos theory and fractal behaviour have been reported in the last few years \cite{gieraltowski2012multiscale,goldberger1996non,hampson2011multifractal,bryce2012revisiting,chowdhury2013surface,hakonen2015current}. 
Physiological signals are highly complex, therefore require an appropriate analysis, which will 
be able to bring us closer to the understanding of the true nature of the process (or processes) 
behind the signal. Traditional analysis, mainly based on the conventional statistical tests of mean, 
median etc. may not be sufficient (see for instance the series of articles of analysis of the cardiac 
rhythm \cite{Goldberger1990chaos,makowiec2006long}), as the important information embedded in the 
signal could be easily lost. Additionally, for the description of the neuromuscular activation 
during functional movements, the influence of the neighbouring muscles due to the location of 
the electrodes over the group of muscles and the modification of the source position in relation 
to each electrodes are the main difficulties with the data interpretation 
\cite{konrad2005abc}. In the discussed case the MultiFractal Detrended Fluctuation Analysis 
(MFDFA) was applied. The proposed method is based on scaling properties of fluctuations in the 
time series. MFDFA developed by Kantelhardt et.al.
\cite{kantelhardt2002multifractal,kantelhardt2009fractal} became a popular method for the wide 
range of application for the study of the nonlinear phenomena \cite{goldberger1996non}. This 
includes aspects of the biomedical signal analysis \cite{gupta1997fractal,ivanov1999multifractality} 
which is the case presented here. The paper is organised as follows: Section 2 describes the 
experimental method. In Section 3 the MFDFA method for data analysis is introduced. The results 
are presented in Section 4. The last Section summarises the results and draws conclusions. 

\section{Method}
\label{method}
\subsection{Subjects and task}
\label{task}
\begin{figure}[htbp]
\centering
\includegraphics[width=.7\textwidth]{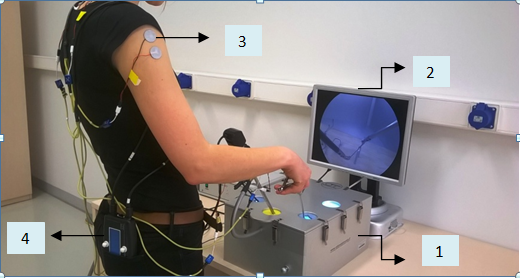} 
\caption{\small{(color online) A volunteer working on a laparoscopic trainer; 1--laparoscopic box, 2--preview of the performed task, 3--pair of electrodes, 4--the measuring device.}}
\label{fig1}
\end{figure}

Six volunteers (equal gender distribution or just 3 male and 3 female), 24--27 years of age,
at similar physical conditions were recruited for the experiment. The experiment was 
conducted on a laparoscopic trainer -- see Fig. \ref{fig1}. All participants were 
right--handed. Novice users had to tie surgical knots using intra--corporeal, double 
handed technique. The knots were tied on the metal half--rings using 15 cm long, 
surgical thread. The task was to tie the largest possible numbers of knots in the 
allotted time. There were two measurement points: First one at the beginning of 
the experiment and the second one after a series of training events. The average 
time required to learn the proper technique took around 3 hours (three 
series of training, \textasciitilde 60 minutes each).

\subsection{Experimental data}\label{data}
To quantify the results of the experiment the surface electromyography method (sEMG) was chosen.  
Muscular activity was recorded from the four groups of muscles on each arm. The selected groups 
potentially exhibit the largest level of an excessive tension and constitute the main muscle sets 
engaged in the performed task. The electrodes were located around separated groups of muscles, namely trapezius 
ridge; deltoids; forearm--long palmar muscle and ulnar wrist flexor; thenar eminence--abductor muscle of thumb and flexor brevis.  We used bipolar concentric surface AgCl electrodes, 15x15 mm in size, with a concentrating 
connector. The inter--electrode distance was set to $1$ cm. Concentric electrodes were used in order 
to compensate the difficulties with proper placement of electrodes in relation to direction of 
muscular fibers and also to compensate the changes which are related to the morphology of the 
shape of the action potential during the movement. 
We took an extra effort to precisely localise the electrodes over the belly of the muscle in order
to avoid the influence from the adjacent muscles and to reduce the cross-talk.
The measurements were conducted with the 8 channel 
surface EMG recorder (OT Bioelettrinica, Torino, IT). The system automatically records the mean value 
of the EMG signal over a time interval of 125 milliseconds. The Average Rectified Value (ARV) measured 
in microvolts was collected in the maximum voluntary contraction mode (MVC). The measurement was 
performed for the complete relaxed state and during the movements before and after the training 
series. For all these states usual recording time took around 45 minutes.

\section{Data analysis}\label{analysis}
The recorded difference of the electric potential present on the skin which in 
turn is related to the action potentials propagating along the muscle fibers 
will be a main source of the analyzed data. As usual the idea was to find 
quantifiers which will allow to justify the actual state of 
the group of muscles. We are mainly concerned by the possibility of comparison 
of the two states not trained and trained ones. The central result of 
multifractal analysis is a multifractal spectrum (mf--spectrum). The complete 
procedure and the detailed step-by-step numerical scheme can be found in 
\cite{kantelhardt2002multifractal,kantelhardt2009fractal,ihlen2012introduction}. 
Here we will only present the general idea of the MFDFA method and describe the parameters 
of interest.

\subsection{MultiFractal spectrum}\label{mfdfa}
The essential aspect of the fractal (and multifractal) analysis lies in the self--similarity.
The time series $x(t)$ observed at a time scale $t$ is said to be statistically self--similar 
with the time series $x(kt)$ observed at $k$ times longer time scale $kt$ when the following
relation is fulfilled
\begin{equation}\label{eq:hurst}
x(kt) \equiv k^H x(t).
\end{equation}
The exponent $H$ characterises the type of self--simmilarity. The relation (\ref{eq:hurst}) 
describes the system for which the magnification of a small part is not 
statistically different from the whole.

The self--similar (self--affine) time series are often express as a fractal, but in less rigorous 
terminology, see for details \cite{kantelhardt2009fractal,Lakhatia1986,Abry.book.2009}. 
The development of methods for estimating self--similarity exponents has enabled a possibility
for the precise description of the the complex multi-scale organization of the signal, 
even if the signal itself cannot be regarded as a fractal in a strict sense \cite{Abry.book.2009}. 
MFDFA is one among several methods which is widely used to calculate a set of similarity exponents. 
This method is based on the analysis of the scaling properties of the signal's fluctuations. 
It offers a scheme to obtain the multifractal spectrum which indicates the frequency of the 
occurrence of the singularities. 

In the following we will introduce the typical MFDFA as presented in \cite{kantelhardt2002multifractal} 
and \cite{ihlen2012introduction}. In short, the analysis requires
the following stages: Suppose that we have time series with $N$ data points
$\{x_i\}_1^N$, we perform than four consecutive steps

\begin{itemize}
\item[(i)] Calculate the profile $y_i$ as the cumulative sum from the data
with the subtracted mean
\begin{equation}\label{eq1}
y_i = \sum_{k=1}^i [x_i - \langle x \rangle].
\end{equation}

\item[(ii)] The cumulative signal is split into $N_s$ equal non-overlapping segments of
size $s$. Here, for the length of the segments we use the power of two, $s = 2^r$.
%where $r = 4, \dots,\left \lfloor \log_2(N/10) \right \rfloor$. 
Typically the exponents $r$ would range from $4$ up to 
$\left \lfloor \log_2(N/10) \right \rfloor$. However, the minimum sample size must be 
larger than the polynomial order to prevent over-fitting of polynomial
trend. The use of $N/10$ for the upper limit means that at least $10$ segments will be used
in calculations. Larger segment sizes will result with rather weak statistics.
Usually the length of the data will not be accordant
with the power of two and some data parts would have to be dropped from the analysis.
Therefore the same procedure should be performed starting from the last index, and
in turn the $2 N_s$ segments will be taken into account.

\item[(iii)] Calculate the local trend $y^m_{v,i}$ for $v^{th}$ segment by means of the
least--square fit of order $m$. Then determine the variance
\begin{equation} 
F^2(s,v) \equiv {1 \over s} \sum_{i=1}^{s} \left( y^m_{v,i} - y_{v,i} \right)^2 \label{fsdef} 
\end{equation}
for each segment $v = 1, \ldots, N_s$. The same procedure has to be 
repeated in the reversed order (starting from the last index).
Next determine the fluctuation function being
the $q^{th}$ statistical moment of the calculated variance.
\begin{eqnarray}
\label{eq2}
F_q(s) &=& \left(\frac{1}{2N_s}\sum_{v=1}^{2N_s} [F^{2}(s,v)]^\frac{q}{2} \right)^\frac{1}{q},
\quad q \ne 0,\\
F_0(s) &=& \exp \left\{ {1 \over 4 N_s}
\sum_{\nu=1}^{2 N_s} \ln \left[F^2(s,\nu)\right] \right\},
\quad q = 0.
\end{eqnarray}
The above function needs to be calculated for all segment sizes $s$.
We have exploited several different orders of the fitted polynomials and end up with no
statistical difference between the results. Here we will present the analysis with
the quadratic fit.

\item[(iv)] In the last step the determination of the scaling
law of the fluctuation function (\ref{eq2}) is performed by means of the log--log
plots of $F_q(s)$ versus segment sizes $s$ for all values of $q$. The function
$F_q(s) \sim s^{h(q)}$ is naturally smaller for the smaller fluctuations, which
results in the increasing function with the increasing segment size.
From the $h(q)$ called generalized Hurst exponent we are able to determine several quantifies.
Firstly, we work out the mass exponent using the formula
\begin{equation}\label{mass}
\tau(q) = q h(q) - 1.
\end{equation}
The mass exponent $\tau(q)$ it is used to calculate a $q$--order 
singularity exponent $\alpha = \tau'(q)$. This quantity is also known as a H\"older exponent.
From the above the $q$--order singularity dimension 
\begin{equation}\label{spectrum}
D(q) = q \alpha - \tau(q) = q[\alpha - h(q)] + 1.
\end{equation}
can be constructed.
The singularity dimension $D(q)$ is related to the mass exponent $\tau(q)$ by Legendre transform.
\end{itemize}

\begin{figure}[htbp]
\centering
\includegraphics[scale=.8]{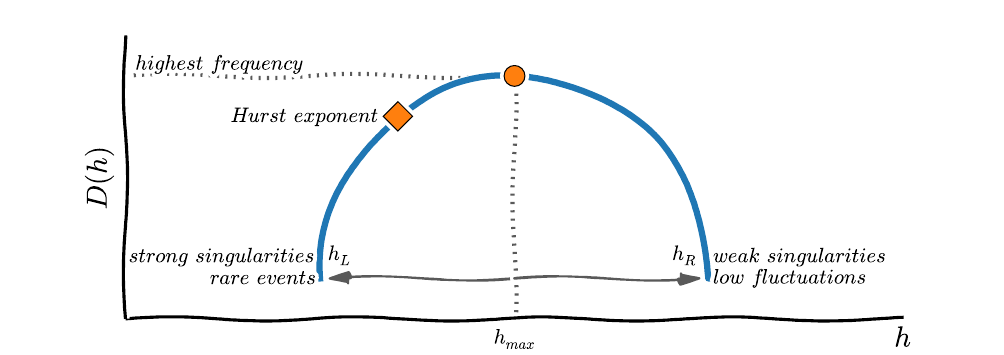} 
\caption{(color online) \small{Schematic picture of the multifractal spectrum with 
the characteristic parameters: spectrum width, Hurst exponent and 
$h_{max}$ (the most probable singularity)}}
\label{fig2}
\end{figure}

The multifractal spectrum shown schematically in Fig. \ref{fig2} identifies the deviation 
of the fractal structure within the time periods for large and small fluctuation 
\cite{ihlen2012introduction}. The rare events are defined as the smallest values of 
generalized Hurst exponent $h(q)$ located at the left end of the spectrum.
In the quantitative description of the spectrum we would like to concentrate on the 
spectrum width $\Delta = h_R - h_L$ and the global Hurst exponent $H$ (or self--similarity exponent).

These values describe crucial properties for the recorded signals. The spectrum width 
determines the diversity of periods with the different scales (for high and low fluctuation). 
The values of Hurst exponent describe the nature of noise found 
in the time series. The range of the Hurst exponent values can be interpreted 
\cite{kantelhardt2002multifractal,ihlen2012introduction,gieraltowski2012multiscale}
as follows 
$h \in (0,0.5)$ indicates the antipercistency of the signals, 
$h=0.5$ represent an uncorrelated noise, 
$h\in(0.5,1)$ indicates the persistency of the series.
This interpretation of the entire signal is valid only if the data exhibits the
monofractal character. For the multifractal systems the set of the exponents
is needed which is caused by the local character of the fluctuations.
The shape of mf-spectrum itself has wide, meaningful interpretation \cite{makowiec2006long,ihlen2012introduction}. 
%In this paper we would like to focus on the quantitative comparison of 
%the two mf-spectrum parameters, the global Hurst exponent $H$ and the spectrum width 
%$\Delta$, for different states of the muscle \textit{modus operandi}.

\begin{figure}[t]
\centering
\includegraphics[scale=1]{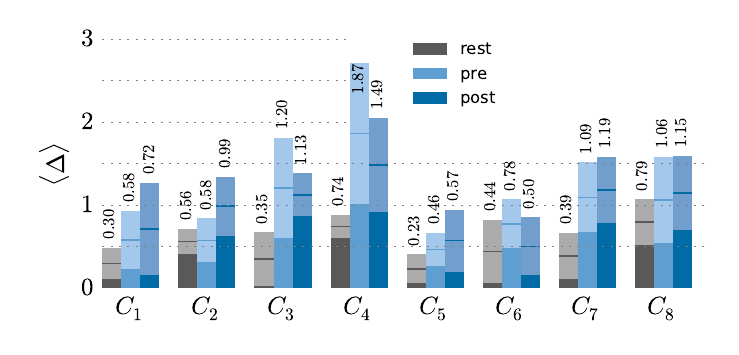} 
\includegraphics[scale=1]{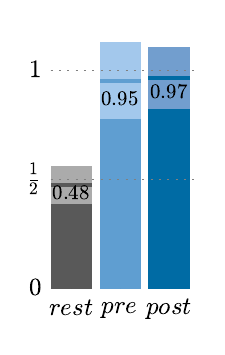} 
\caption{\small{(color online) Mean values of the spectral width $\Delta$. Left panel: comparison of 
mean values for the individual groups of muscles (see text for details). Right panel: comparison of 
the mean values for three different states -- at rest, before the training (pre) and after the 
training (post). Numbers refer to the actual values of means. The confidence interval is
depicted with the slightly lightened colors for each bar separately.}}
\label{fig3}

\end{figure}

\section{Results}\label{results}
Despite many advantages offered by MFDFA method in the application to the complex biomedical data, 
some particular steps require from the users the individual decisions which can have significant 
impact on the final results. The main issue concerns the choice of the scaling range $s$ for the 
proper estimation of the fluctuation function $F_q(s)$ (\ref{eq2}). In the literature one can 
find some useful advices for the appropriate selection the range of the scales 
\cite{gieraltowski2012multiscale,ihlen2012introduction}. 

\begin{figure}[ht!] 
\begin{center} 
\subfigure[$F_q(s)$ before training]{ 
\includegraphics[width=0.45\textwidth]{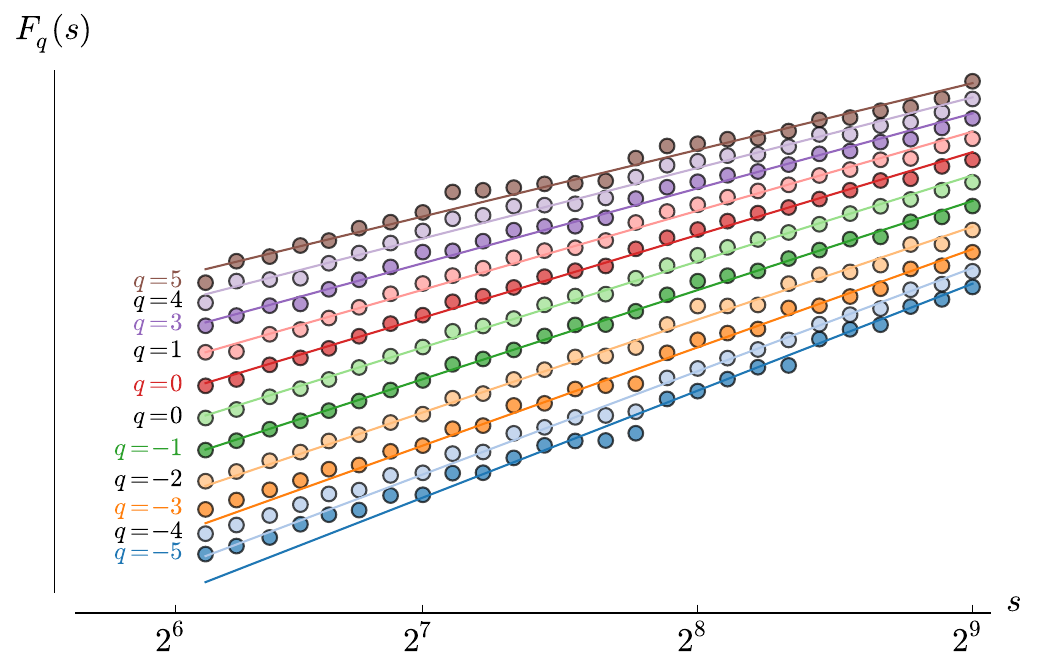} 
\label{fig81} 
}
\subfigure[$F_q(s)$ after training]{ 
\includegraphics[width=0.45\textwidth]{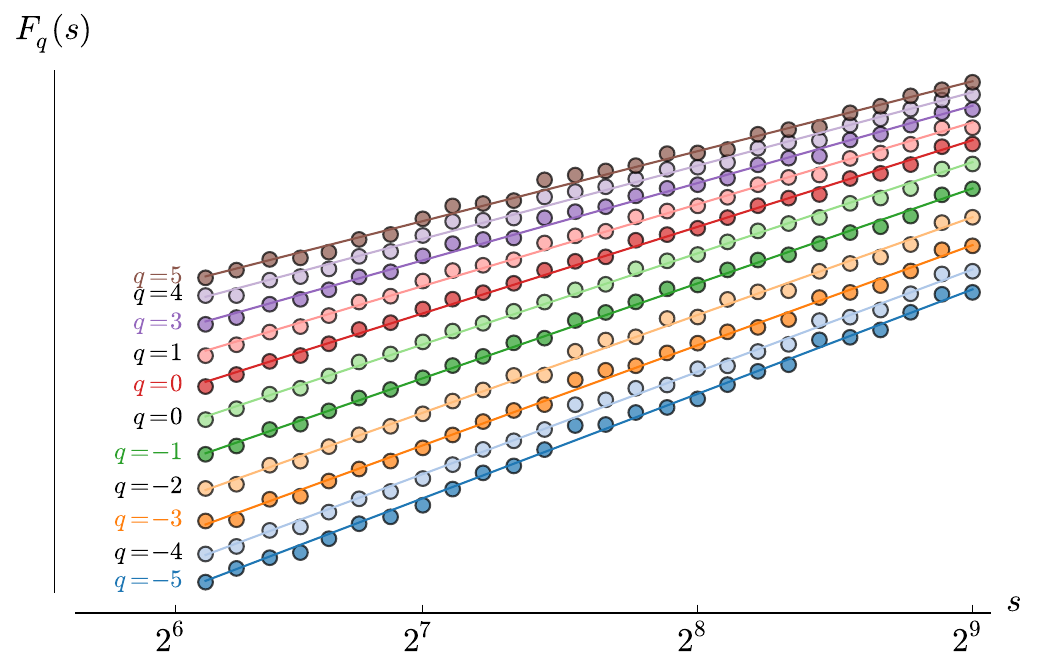}
\label{fig82} 
}
\subfigure[mf--spectrum before training]{ 
\includegraphics[width=0.45\textwidth]{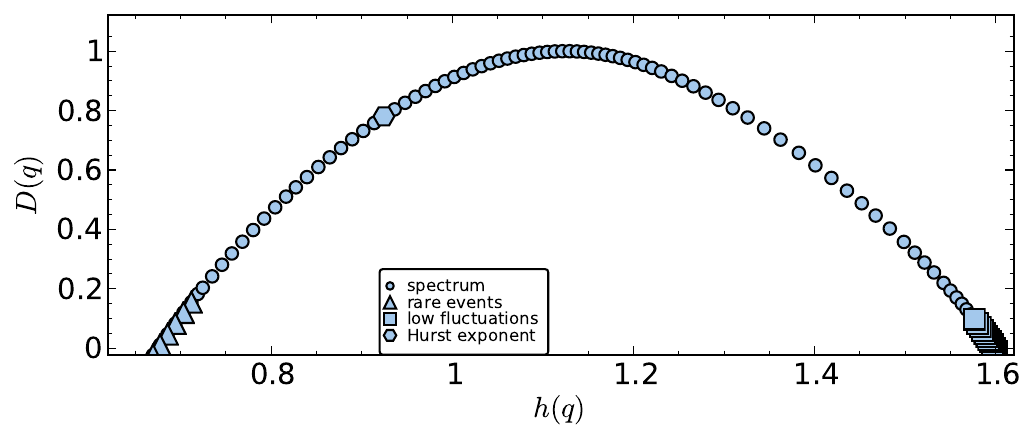}
\label{fig83} 
}
\subfigure[mf--spectrum after training ]{ 
\includegraphics[width=0.45\textwidth]{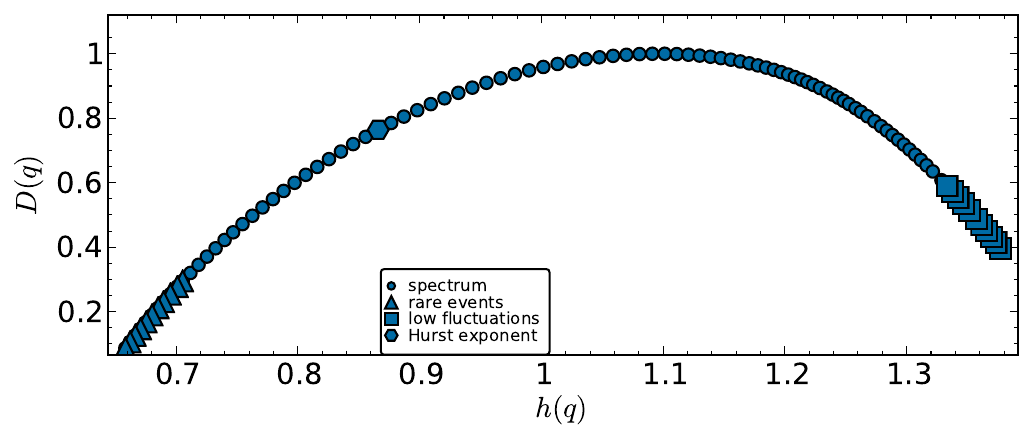}
\label{fig84} 
}
\caption{The example of double logarithmic dependence $F_q(s)$ vs $s$ together with the corresponding multifractal spectra presented for the forearm muscle group from the right hand ($C_6$). } 
\label{fig8} 
\end{center} 
\end{figure}

The length of the analysed time series $N$ consists of around $21000$ data points. For the calculations 
presented in this work the scales (segments length) $s \in (64,512)$ and typical $q\in (-5,5)$ 
were chosen. The examples of the 
double logarithmic dependence $F_q(s)$ vs $s$ together with the corresponding multifractal spectra are 
presented in the Fig. \ref{fig8}.  
Within the selected range of scales the linear dependence of the fluctuation function $F_q(s)$ on the
segment length $s$ can be observed. Also it can be noticed that for the presented working states 
(before and after the training) the spectrum is relatively wide and there is no significant difference 
between the values of mf--spectrum width before and after training.

%\section{Statistics}
\subsection{Spectrum width}\label{sec:spectrum}
The mean values of the spectral width are presented  in the Tab. \ref{tab:width}. 
Channels are assigned 
to each group of muscles. $C_1$--$C_4$ represent the left arm 
and $C_5$--$C_8$  indicate  muscles from the right arm. 
$C_1$ and $C_5$ form a pair of the respective trapezius ridge group of muscles located on
the left and right arms. Similarly the channels $C_2$ and $C_6$ record signals from the deltoids, 
$C_3$ and $C_7$ from the forearm muscle group (long palmar muscle and ulnar wrist flexor), and 
finally $C_4$ and $C_8$ from the group of thenar muscles. 
\begin{table*}
\centering
\caption{Mean values of the spectrum width for all channels and at each state.}
\label{tab:width}
\begin{tabular}{|c|c|c|c|c|c|c|c|c|}
\hline
\multicolumn{5}{|c|}{Average spectrum width $\langle \Delta \rangle$}                               \\ \hline
Channel number  & $C_1$  & $C_2$  & $C_3$  & $C_4$  & $C_5$  & $C_6$  & $C_7$  & $C_8$  \\ \hline
Relaxation      & \begin{tabular}[c]{@{}c@{}}0.3\\ $\pm 0.074$\end{tabular} & \begin{tabular}[c]{@{}c@{}}0.564\\ $\pm 0.058$\end{tabular} & \begin{tabular}[c]{@{}c@{}}0.35\\ $\pm 0.13$\end{tabular} & \begin{tabular}[c]{@{}c@{}}0.743\\ $\pm 0.054$\end{tabular} & \begin{tabular}[c]{@{}c@{}}0.233\\ $\pm 0.068$\end{tabular} & \begin{tabular}[c]{@{}c@{}}0.441\\ $\pm 0.147$\end{tabular} & \begin{tabular}[c]{@{}c@{}}0.39\\ $\pm 0.11$\end{tabular}  & \begin{tabular}[c]{@{}c@{}}0.794\\ $\pm 0.108$\end{tabular} \\ \hline
Before training & \begin{tabular}[c]{@{}c@{}}0.58\\ $\pm 0.14$\end{tabular} & \begin{tabular}[c]{@{}c@{}}0.577\\ $\pm 0.103$\end{tabular} & \begin{tabular}[c]{@{}c@{}}1.204\\ $\pm 0.235$\end{tabular} & \begin{tabular}[c]{@{}c@{}}1.865\\ $\pm 0.33$\end{tabular} & \begin{tabular}[c]{@{}c@{}}0.465\\ $\pm 0.078$\end{tabular} & \begin{tabular}[c]{@{}c@{}}0.778\\ $\pm 0.114$\end{tabular} & \begin{tabular}[c]{@{}c@{}}1.095\\ $\pm 0.164$\end{tabular}  & \begin{tabular}[c]{@{}c@{}}1.063\\ $\pm 0.2$\end{tabular} \\ \hline
After training & \begin{tabular}[c]{@{}c@{}}0.716\\ $\pm 0.216$\end{tabular} & \begin{tabular}[c]{@{}c@{}}0.987\\ $\pm 0.14$\end{tabular} & \begin{tabular}[c]{@{}c@{}}1.126\\ $\pm 0.1$\end{tabular} & \begin{tabular}[c]{@{}c@{}}1.488\\ $\pm 0.22$\end{tabular} & \begin{tabular}[c]{@{}c@{}}0.574\\ $\pm 0.15$\end{tabular} & \begin{tabular}[c]{@{}c@{}}0.504\\ $\pm 0.14$\end{tabular} & \begin{tabular}[c]{@{}c@{}}1.187\\ $\pm 0.155$\end{tabular}  & \begin{tabular}[c]{@{}c@{}}1.147\\ $\pm 0.174$\end{tabular} \\ \hline
\end{tabular}
\end{table*}

The smallest value of the spectrum width for each channel occurs at 
the relaxation state. The Wilcoxon test at the significance level of $\alpha=0.05$ was 
used in order to compare the relaxation state with the work states before and 
after the training. With the exception of $C_5$ and $C_8$ all other channels indicate 
a statistical significance at the selected level ($p<0.05$). This effect is clearly apparent
in the right panel of Fig. \ref{fig3} where the mean values of the mf--spectrum 
$\langle \Delta \rangle$ 
calculated from all the channels in the three states (rest, pre, post) are presented. 
The left diagram in Fig. 
\ref{fig3} suggests the similar character of the spectrum width for the 
corresponding muscles in the right and left arms. 
For the series obtained before and after the training the highest value of the 
mf--spectrum width occurs for the groups of the forearm ($C_3$ and $C_7$) and thenar 
($C_4$ and $C_8$) muscles, see Tab. \ref{tab:width} for details. The spectrum width 
can serve as an effective predictor for identification of the relaxation state of
the muscle activity. On the other hand the width $\Delta$ alone cannot distinguish
between the working states before and after the training. 

\subsection{Hurst exponent}\label{sec:hurst}
\begin{figure}[ht!]
\centering
\includegraphics[scale=1]{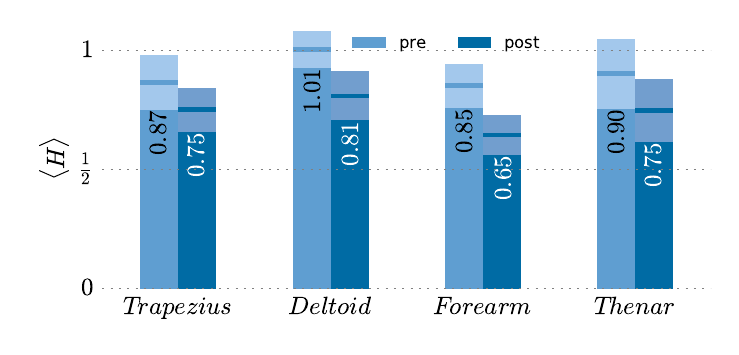}
\includegraphics[scale=1]{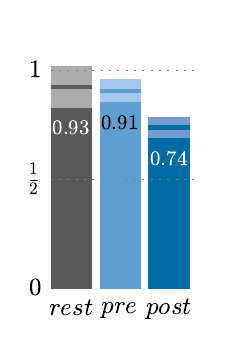} 
\caption{\small{(color online)
Mean values of the Hurst exponent. Left panel: comparison of mean values for the individual groups of 
muscles (see text for details). Right panel: comparison of the mean values for three different states 
-- at rest, before the training (pre) and after the training (post). Bars indicate standard 
deviations. Numbers refer to the actual values of means. The confidence interval is depicted
with the slightly lightened colors for each bar separately.}}
\label{fig4}
\end{figure}

\begin{figure}[ht!]
\centering
\includegraphics[scale=1]{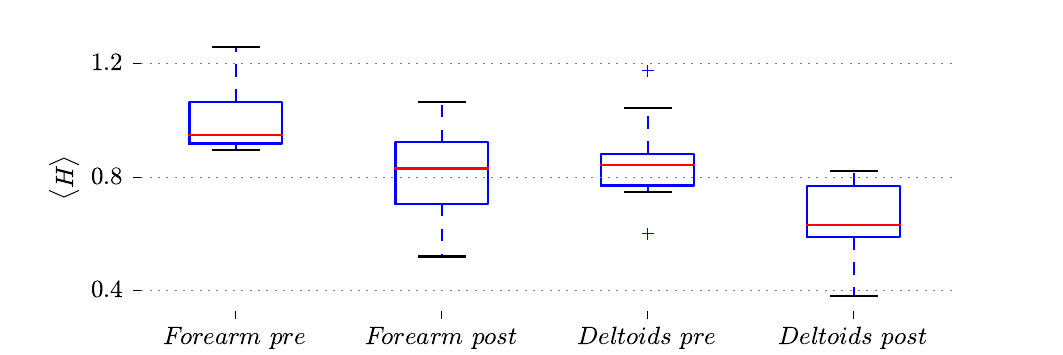} 
\caption{\small{(color online)
The box-and-whisker diagram of the average Hurst exponent $\langle H \rangle$ for two different
groups of muscles at two different working states. The average Hurst exponents were calculated 
for both arms together. The plot for the forearm muscle group 
(long palmar muscle and ulnar wrist flexor) is shown on two diagrams on the l.h.s. The two 
diagrams on the r.h.s correspond to the deltoids. \textit{Pre} and \textit{post} reflects 
the abilities of using the laparoscopic tools before and after the training, respectively.
The boxes correspond to the estimated quartiles and whiskers indicate a variability outside 
the upper and lower quartiles, i.e. the minimum and maximum values.}}
\label{fig5}
\end{figure}

The most important aspect of this work was the identification of the statistically different 
parameters which could be applied to distinguish between the signals obtained from the skilful and
inexperienced student. This quantifiers could be than use for automatic evaluation of 
one's ability of handling laparoscopic tools.
Firstly we would like to focus on these two working states (pre and post) by means 
of the Hurst exponent $H$. 
In contrast to the just analysed width $\langle \Delta \rangle$, the mean Hurst exponent 
$\langle H \rangle$ 
indicates the statistical difference between of the data recorded before and after the 
training, c.f. Fig. \ref{fig4}. 
We grouped together the values for the corresponding clusters of muscles from the left 
and right arm for all six volunteers and calculated the arithmetic mean values of the Hurst 
exponent $\langle H \rangle$ for the corresponding channels separately. This values are 
summarized in the Table \ref{tab:hurst} and presented in the left panel of Fig. \ref{fig4}.
For all the muscle groups the values 
of the Hurst exponent of the series after the training (post state) are significantly lower, 
see Fig. \ref{fig4}. The statistical significance occurs for deltoids ($C_2$ and $C_6$, with $p=0.0096$) and 
forearm ($C_3$ and $C_7$, with $p=0.0077$). The visual representation on this 
dependence is presented as the box plot in Fig. \ref{fig5}.

\begin{table}[hb!]
\caption{Average values of the Hurst exponent $\langle H \rangle$ for the 
corresponding groups of muscles for the left ($C_1$--$C_4$) and right ($C_5$--$C_8$) 
hand at each of the three recorded states.}
\label{tab:hurst}
\centering
\begin{tabular}{|l|c|c|c|c|}
\hline

\multicolumn{5}{|c|}{Average Hurst exponent $\langle H \rangle$} \\ \hline
Muscles group & Trapezius ridge & Deltoid & Forearm & Thenar \\
{} & ($C_1+C_5$) & ($C_2+C_6$) & ($C_3+C_7$) & ($C_4+C_8$) \\ \hline
Relaxation/rest & $0.791 \pm 0.071$ & $0.894 \pm 0.097$ & $0.715 \pm 0.055$ & $1.304 \pm 0.057$ \\ \hline
Before training/pre & $0.868 \pm 0.053$ & $1.006 \pm 0.035$ & $0.853 \pm 0.043$ & $0.904 \pm 0.067$ \\ \hline
After training/post & $0.752 \pm 0.042$ & $0.811 \pm 0.047$ & $0.646 \pm 0.0376$ & $0.749 \pm 0.06$ \\ \hline
\end{tabular}
\end{table}

\subsection{Thenar group of muscles}\label{sec:thenar}
Upon the comparison of the signals acquired at rest and before the training sessions 
with those after the training the significant difference of the calculated averages of the Hurst 
exponents can be seen. This result is 
demonstrated in the right panel of the Fig. \ref{fig4}, where the mean values 
of Hurst exponent taken from the all the muscles together are presented for each of the states
rest (gray), pre-- (light blue) and post--training (blue). 
The dominant impact of this dissimilarity lies in the mean value of the Hurst exponent of 
the thenar muscles which show a statistically significant ($p=0.0037$) increase of the exponent 
for the relaxation state separately. 
This findings were quite unexpected as they indicate the different character of irregular
component for the rest state of the discussed group of muscles alone.
At the rest state the value is relatively
high $\langle H \rangle_{C_4, C_5}^{rest} = 1.3037$ and is
typically assigned to the integrated noise (random walk) \cite{ihlen2012introduction}. For all the other 
muscle groups, the relaxation states show the persistence character \cite{ihlen2012introduction},
which is characteristic to the noisy nature of the time series, c.f. Tab. \ref{tab:hurst}.
\begin{figure}[ht!]
\centering
\includegraphics[scale=1]{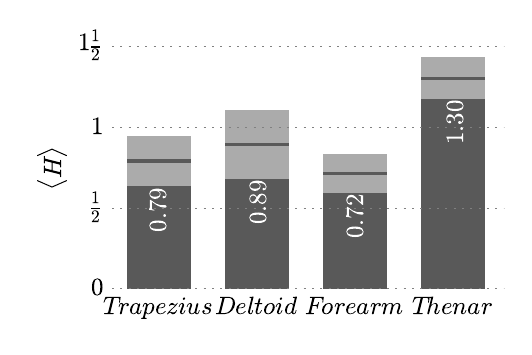}
\includegraphics[scale=1]{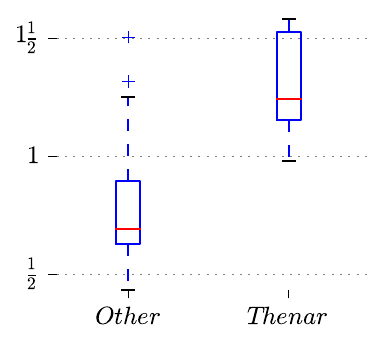}
\caption{\small{(color online)
Average Hurst exponent presented for all group of muscles calculated for both arms together. 
The left panel shows bar chart for all of the groups separately. Please note that the thenar 
group of muscles possesses the highest, Brownian--motion--like, value of $\langle H \rangle$.
All other groups can be interpreted as noisy signals. The confidence interval is
depicted with the light grey for each bar separately.
On the r.h.s. the box-and-whisker diagram of the average Hurst exponent is compared for 
joined groups of the trapezius, deltoid, and forearm muscles \textit{versus} the thenar group.
The boxes correspond to the estimated quartiles and whiskers indicate a variability outside 
the upper and lower quartiles, i.e. the minimum and maximum values.
}}
\label{fig6}
\end{figure}

\begin{figure}[htbp]
\centering
\includegraphics[scale=1]{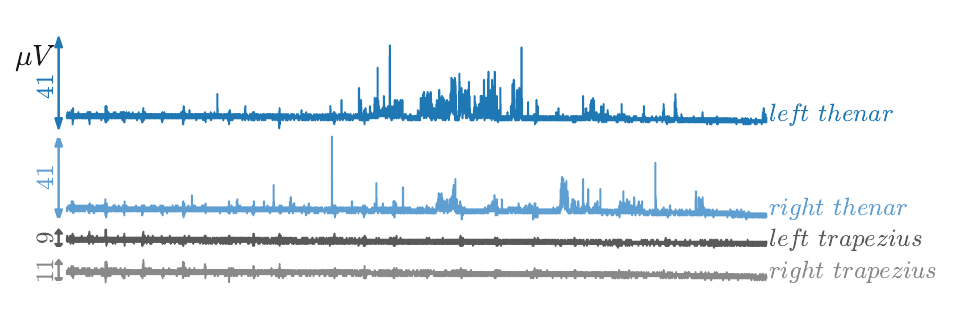}
\caption{\small{(color online)
Exemplary raw sEMG signals acquired from one of the volunteers before the analysis.
The arrows present the ranges of the recorded values of the signal in $\mu$V. 
Groups of muscles from top to bottom: left thenar, right thenar, left trapezius, right
trapezius. The strong difference in the character of the signals can be notice just 
by the naked eye.
}}
\label{fig7}
\end{figure}

The discussed cases are depicted in the Fig. 7 which compares the individual raw signals 
that indicate the biggest difference of the Hurst exponent at the relaxation state.
Fig. \ref{fig7} presents raw time series acquired from one of the volunteers from
four groups of muscles left thenar, right thenar, left trapezius, right
trapezius (top to bottom). The difference in the muscle activity -- significantly
lower amplitudes for the trapezius ridge and much larger fluctuations for the thenar 
group are clearly visible.
In addition, one can notice that for the thenar group the average spectral widths $\Delta$ 
also take the highest values out of all muscle groups (see Tab \ref{tab:width}) for details).

This very muscle group exhibits several difficulties at the measurement. For instance 
the tools constantly touch the skin and can also touch the electrodes from time to time. 
This might cause 
an excessive sweating which in turn influence the signal. Last issue is the area of the 
electrodes in a relation to the area of the muscles. All of the above matter 
mainly at the work states, i.e. when the volunteer perform the actual task with the
laparoscopic tools and are rather irrelevant for the rest state.
%For the rest state, when all the person is required to do is
%simply sit and relax, the only  factor might be the smaller distance between 
%the electrodes compared to the other groups of muscles.

\section{Conclusions}\label{conclusions}
The signal from the surface electromyography recorded during the complex movements 
on the laparoscopic trainer was analysed. The main goal was to find a statistical difference 
between the signals acquired before (pre) and after (post) the training. In addition
the relaxation (rest) state was considered. In order to quantify the complexity of the
series the selected parameters which characterize a multifractal spectrum was used. 
The study of the spectral width does not allow to determine the differences in the 
pre--post states but is is sufficient to distinguish between rest and work states.
Due to the small sample and also the necessity of using the non-parametric test, the 
statistical power of the test was respectively lower. However the investigation of the 
values of the Hurst exponent indicate that this parameter seems to be a better classifier
between the analysed states (before and after the training) than multifractal spectrum width.  
For all of the examined muscle groups for 
both arms the values of the Hurst exponent appear to be lower after the series of
training, however this difference is statistically significant only for the deltoid and 
forearm muscle groups. 
The rather unexpected findings is the character of the thenar muscles. Both the value of 
mf--spectrum width and the global Hurst exponent are distant from all other groups
at the rest state.
In a summary it can be concluded that surface electromyography indicates a potential 
method for the evaluation of complex dynamics of the action potential of the muscles.
The research on the dynamical properties of the sEMG 
signals is still at early stage where many aspects are waiting to be discovered.

\section*{Acknowledgements}
This work was partially supported by the Polish Ministry of Science and Higher Education (Grant K/ZDS/003962)

\end{document}